\documentclass[final,1p,times]{elsarticle} 
\usepackage{graphicx} 
\usepackage{amssymb} 
\usepackage{amsthm} 
\usepackage{lineno} 

\journal{Nuclear Physics A} 
\begin{document} 

\begin{frontmatter} 


\title{Exploration of jet energy loss via direct $\gamma$-charged particle azimuthal correlation measurements}

\author{A. M. Hamed for the STAR collaboration}

\address{Cyclotron Institute, Texas A\&M University, College Station, TX 77843,
USA}

\begin{abstract} 
The multiplicities of charged particles 
azimuthally associated with direct photons and $\pi^{0}$ 
have been measured 
for Au+Au, p+p, and d+Au
collisions at $\sqrt{s_{NN}}$ = 200 GeV in the STAR experiment.
Charged particles with transverse momentum 0.5 $<$ $p_T^{h^{\pm}}$ $<$ 16 GeV/c for p+p and d+Au, and  
3 $<$ $p_T^{h^{\pm}}$ $<$ 16 GeV/c for Au+Au and pseudorapidity $\mid\eta\mid$ $\leq$ 1.5 in coincidence 
with direct photons and $\pi^{0}$ of high transverse momentum 
8 $<$ $p_T^{\gamma,\pi^{0}}$ $<$ 16 GeV/c at $\mid\eta\mid$ $\leq$ 0.9
have been used for this analysis.
Within the considered range of kinematics, the observed suppressions of the associated yields per direct $\gamma$ 
in central Au+Au relative to 
p+p and d+Au are similar and constant with direct photon fractional energy $z_{T}$ ($z_{T}=p_{T}^{h^{\pm}}/p_{T}^{\gamma}$).
The measured suppressions of the associated yields with direct $\gamma$ are comparable to those with $\pi^{0}$. 
The data are compared to theoretical predictions.
\end{abstract} 

\end{frontmatter} 

In inverse Compton scattering $(q+g \rightarrow q+\gamma)$ and quark annihilation  
$(q+\bar{q} \rightarrow g+\gamma)$ the transverse momentum of the outgoing high-$p_{T}$ $\gamma$ counterbalances the  
transverse momentum of the outgoing parton, modulo negligible corrections from initial-state radiation.
Therefore, direct $\gamma$-hadron azimuthal correlations have been proposed as a good probe for the jet energy loss, especially its 
dependence on the parton initial energy, in the medium created at relativistic heavy ion 
collisions [1]. Moreover, in heavy ion collisions, the produced $\gamma$ departs the medium upon creation 
without any further interactions, and
therefore doesn't 
experience any geometrical bias which might arise from the hard-scattering vertex position 
inside the medium. Accordingly, a comparison of direct
$\gamma$-hadron azimuthal correlations with di-hadron azimuthal correlations could provide 
additional constraints on the path-length dependence of energy loss.   
We report on the results of $\gamma$-charged particles ($\gamma$-ch) and 
$\pi^{0}$-charged particles ($\pi^{0}$-ch) azimuthal correlation measurements in the STAR experiment.

The most challenging aspect of this analysis is to separate the direct photons
from background photons.
In fact the remnants of any source of background could obscure 
the parton initial energy, and cause the $\gamma$-jet coincidence measurements to suffer from the same geometrical biases as
of di-hadron analysis. Among the hadronic decays, $\pi^0$ and $\eta$ are the main 
known sources of direct photon backgrounds in hadronic and nuclear collisions. The STAR experiment at RHIC is very well 
adapted for such challenge through its Barrel Electromagnetic Calorimeter (BEMC) [2]. 
Utilizing the transverse profile technique of electromagnetic showers in the STAR Shower Maximum Detector with 
its fine segmentation ($\Delta\eta\approx 0.007,\Delta\phi\approx 0.007$), 
discrimination between the $(2\gamma$)/$(1\gamma)$ showers up to $p_T^{\pi^{0}}$ $\sim$ 26 GeV/c 
becomes possible. Furthermore, the BEMC tower size ($\Delta\eta=0.05,\Delta\phi=0.05$) is typically greater than the 
angular separation between the two photons resulting from a symmetric decay of high-$p_{T}$ $\pi^{0}$.
Therefore, the determination of $\pi^{0}$'s $p_{T}$, without invariant mass reconstruction of $\pi^{0}$, is attainable.

The STAR detector is very well-suited for azimuthal correlations due to the full coverage in azimuth of the BEMC and the 
STAR main Time Projection Chamber (TPC) [3].
Using a level-2 high-$E_T$ tower trigger to tag $\gamma$-jet events, in 2007 the 
STAR experiment collected an integrated luminosity of 535 $\mu${b}$^{-1}$ of Au+Au collisions at 
$\sqrt{s_{NN}}$ = 200 GeV.
As a reference measurement, the p+p data taken in 2006 with integrated luminosity 
of 11 pb$^{-1}$, and 34 nb$^{-1}$ of d+Au data taken in 2008 at $\sqrt{s_{NN}}$ = 200 GeV
are analyzed. 
In this analysis the $\pi^{0}$/$\gamma$ discrimination was 
carried out by making cuts on the shower shape, where the $\pi^{0}$ 
identification cut was adjusted in order to obtain a direct $\gamma$ free ($\pi^{0}_{rich}$) sample and 
a sample rich in direct $\gamma$ ($\gamma_{rich}$). 
\begin{figure}
\begin{center}
   \resizebox{140mm}{180pt}{\includegraphics{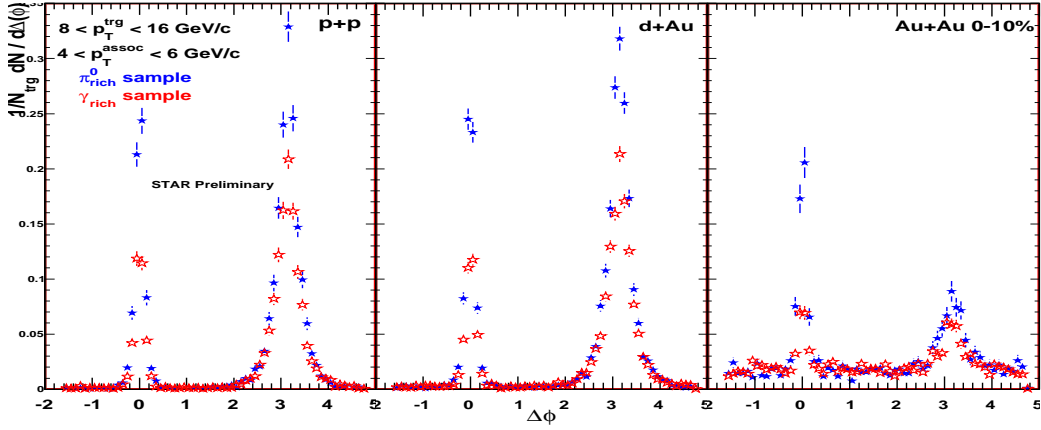}} 
    \caption{Azimuthal correlation histograms of $\gamma_{rich}$ sample 
    and $\pi^{0}_{rich}$ sample 8.0 $<$ $p_T^{\gamma,\pi^{0}}$ $<$ 16.0 GeV/c at $\mid\eta\mid$ $\leq$ 0.9 
    with associated charged particles 4.0 $<$ $p_T^{h^{\pm}}$ $<$ 6.0 GeV/c at $\mid\eta\mid$ $\leq$ 1.5 
    for p+p, d+Au and 0-10$\%$ Au+Au collisions at
    $\sqrt{s_{NN}}$ = 200 GeV.}
      \end{center} 
\end{figure} 

Figure 1  
shows the azimuthal correlation for $\gamma_{rich}$ sample triggers and $\pi^{0}_{rich}$ triggers for p+p, d+Au, and Au+Au
0-10$\%$. 
The medium effect is obviously seen in the away side 
as it was previously reported [4]. 
As predicted, the $\gamma_{rich}$ 
triggered sample has lower near-side yields than 
these of the $\pi^{0}_{rich}$ but not zero. The non-zero near-side yield for the $\gamma_{rich}$ sample is expected due 
to the remaining contributions of the widely separated photons from other sources. 
It is noticeable that the ratio of the away side to near side yields of the $\gamma_{rich}$ sample is slightly greater than that 
of $\pi^{0}_{rich}$ sample which might be due to the effect of direct $\gamma$.

In order to determine the combinatorial background level, the relative 
azimuthal angular distribution of the associated 
particles with respect to the trigger particle is simply fitted with two Gaussian peaks and a straight line.  
The near- and away-side yields, $Y^{n}$ and $Y^{a}$, of associated particles per trigger are extracted by 
integrating the $\mathrm 1/N_{trig} dN/d(\Delta\phi)$ distributions in $\mid\Delta\phi\mid$ $\leq$  0.63 
and $\mid\Delta\phi -\pi\mid$  $\leq$  0.63 respectively. 
Although as discussed later, the purity of the $\pi^{0}_{rich}$ sample is not 
critical for this analysis as long as 
$\pi^{0}_{rich}$ sample is free of direct $\gamma$, a general agreement between the results from this 
analysis ($\pi^{0}$-ch) and the previous analysis (ch-ch) [4] is clearly seen in Fig. 2.(left), which shows  
the $z_{T}= p_{T}^{assoc}/p_{T}^{trig}$
dependence of the associated yield normalized per trigger $D(z_{T}$). 
\begin{figure}
\begin{center}
\begin{tabular}{cc}
   \resizebox{65mm}{215pt}{\includegraphics{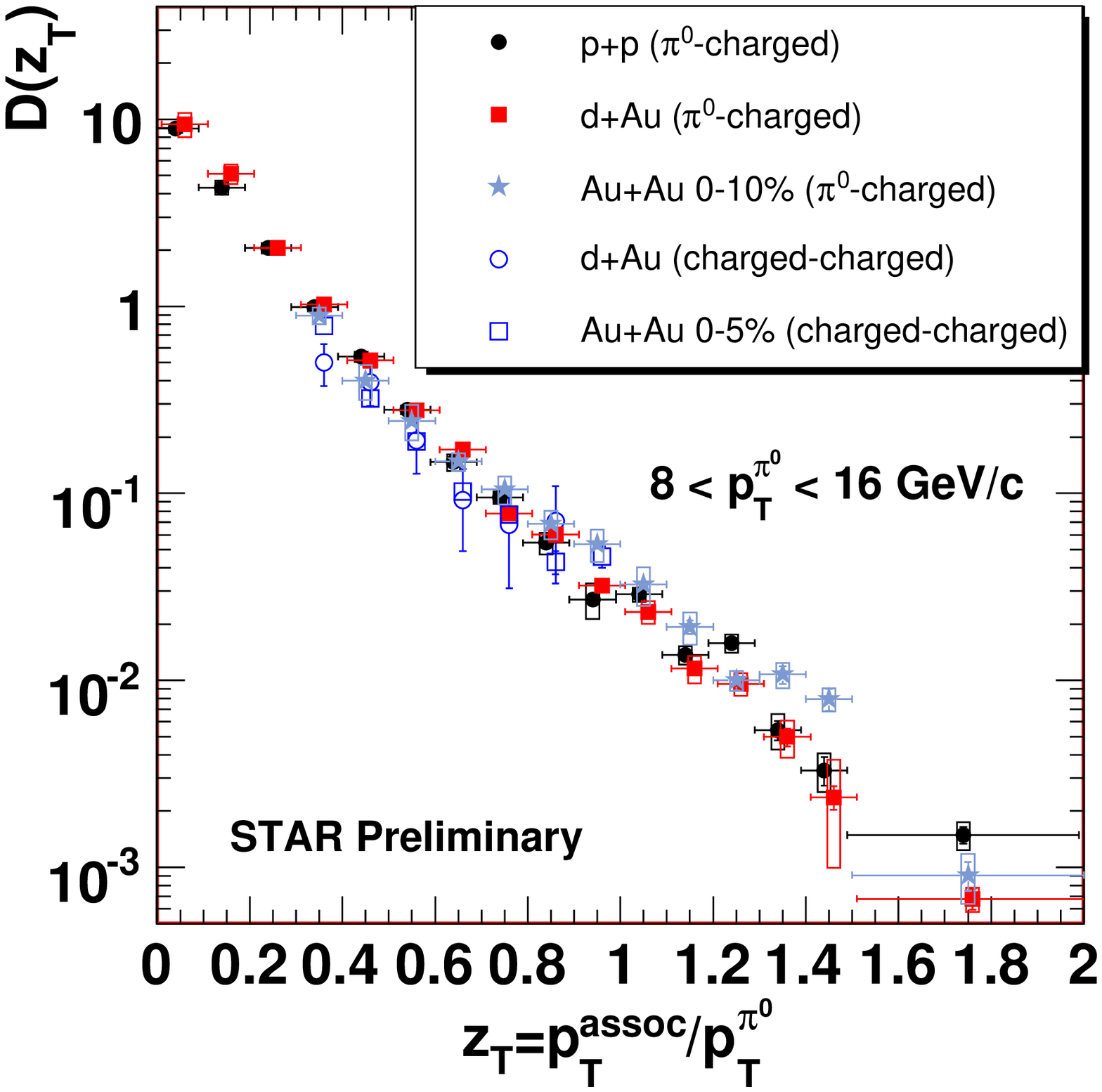}}
   \resizebox{65mm}{200pt}{\includegraphics{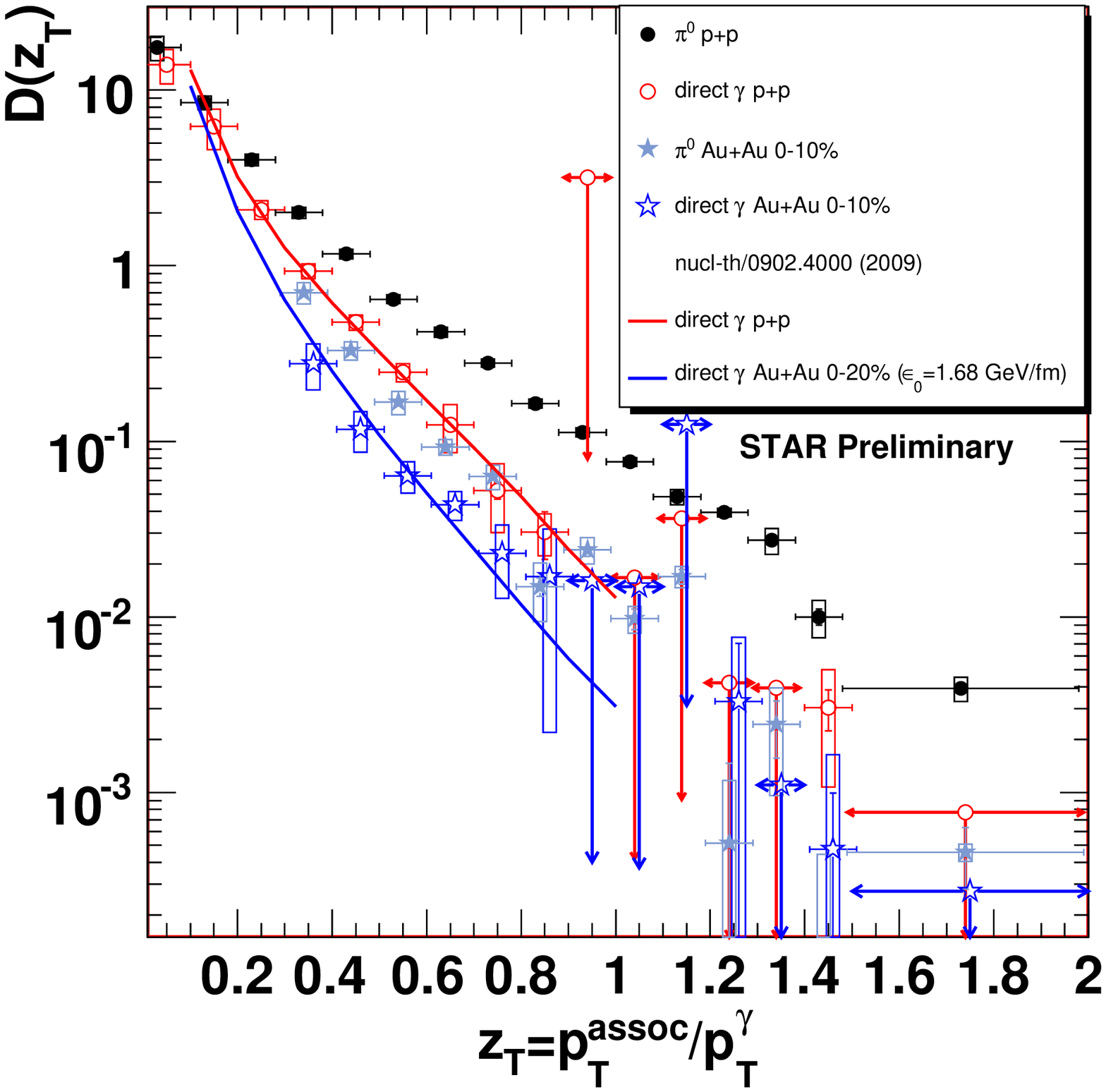}} \\
           \end{tabular}
    \caption{(Left panel) The associated near side yield normalized per hadron trigger, the ch-ch data (open symbols) are from [4].  
             (Right panel) The associated away side yield normalized per trigger for direct $\gamma $ (open symbols) and $\pi^{0}$  
	     (closed symbols) of 8.0 $<$ $p_T^{\gamma,\pi^{0}}$ $<$ 16.0 GeV/c for p+p, and 0-10$\%$ Au+Au collisions at
    $\sqrt{s_{NN}}$ = 200 GeV.
	     The upper limits are for 90$\%$ confidence levels. Theoretical curves are within NLO calculations and $\epsilon_{0}$ is the
	     energy loss parameter proportional to initial average gluon density $\rho_{0}$ [5].}
      \end{center} 
\end{figure}
\begin{figure}
\begin{center}
\begin{tabular}{cc}
   \resizebox{70mm}{200pt}{\includegraphics{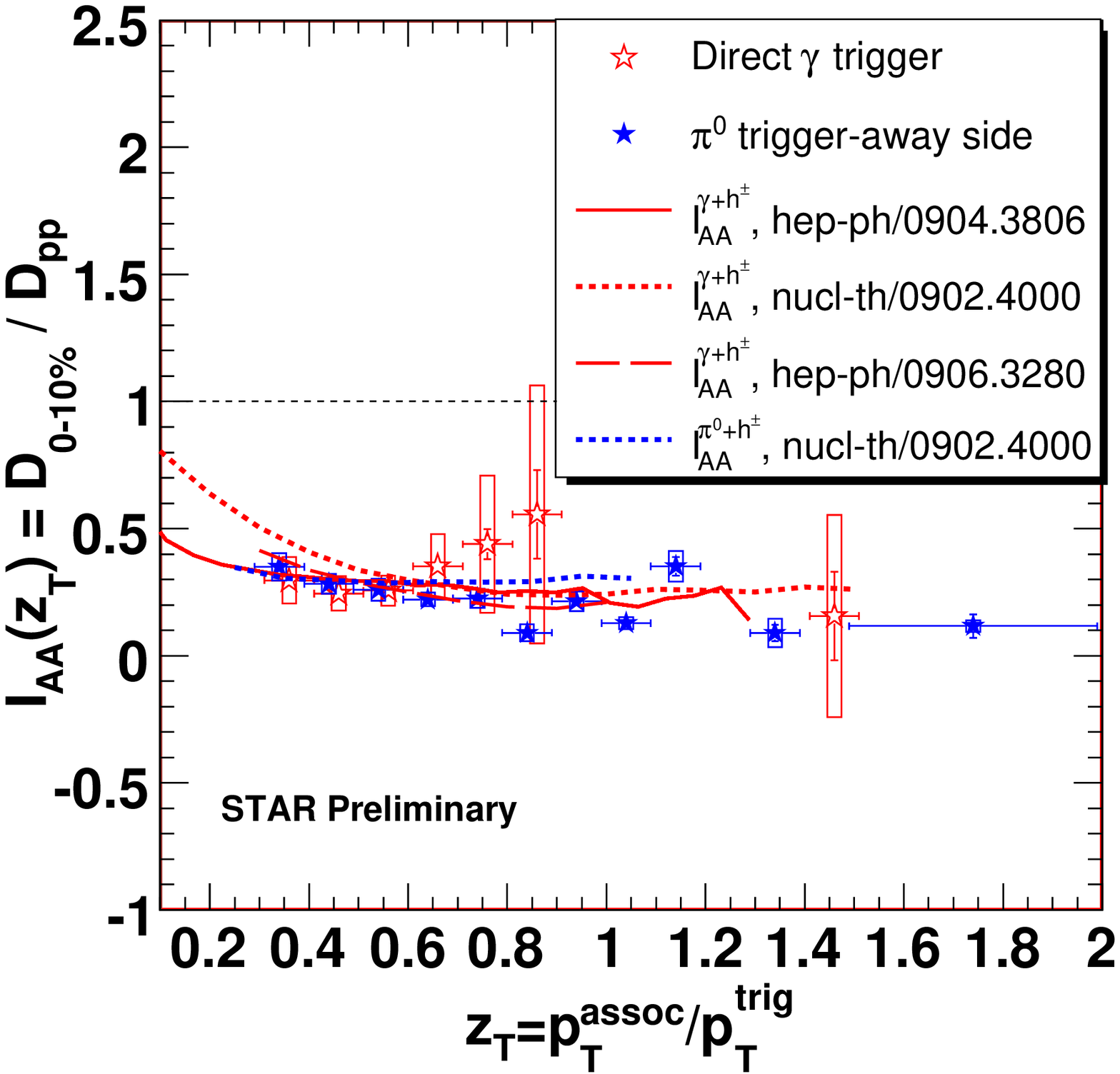}} 
   \resizebox{70mm}{200pt}{\includegraphics{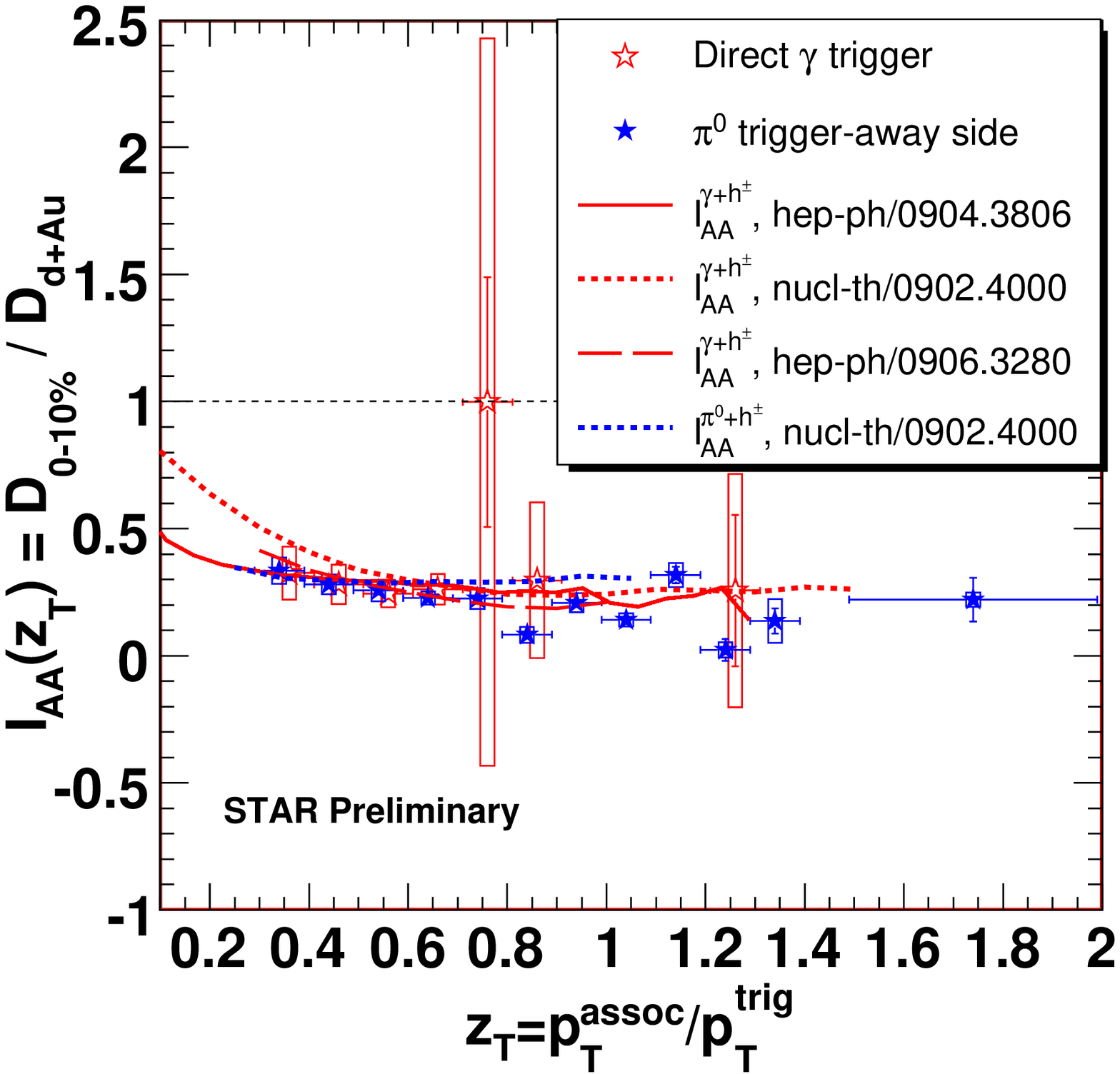}} \\
        \end{tabular}
    \caption{Ratio $I_{AA}$ of $\pi^{0}$ (closed symbols) and direct $\gamma$ (open symbols) of 8.0 $<$ $p_T^{\gamma,\pi^{0}}$ $<$ 16.0 GeV/c at $\sqrt{s_{NN}}$ = 200
    GeV relative to p+p (left panel) and relative to d+Au (right panel). Theoretical predictions are from [5,6,7]. 
    Reference [5] considers the NLO contributions, reference [6]
     examines schematic conjectures of energy loss, and reference [7] incorporates the complete set of photon production channels.}
      \end{center} 
\end{figure}

Assuming zero near-side yield for direct photon triggers and a  
sample of $\pi^{0}$ which is free of direct $\gamma$, the away-side yield of hadrons correlated with the direct photon is 
extracted as
\begin{equation}
Y_{\gamma_{direct}+h}=\frac{Y^{a}_{\gamma_{rich}+h}-{\cal{R}} Y^{n}_{\gamma_{rich}+h}}{1-r} \label{eq:one}
\end{equation}
where ${\cal{R}}={Y^{a}_{\pi^{0}_{rich}+h}}/{Y^{n}_{\pi^{0}_{rich}+h}}$, $r={Y^{n}_{\gamma_{rich}+h}}/{Y^{n}_{\pi^{0}_{rich}+h}}$, 
$1-r={{N^{\gamma_{dir}}}}/{N^{\gamma_{rich}}}$,
and $N^{\gamma_{dir}}$ and $N^{\gamma_{rich}}$ are the numbers of triggers of direct $\gamma$ and $\gamma_{rich}$ respectively.
In Eq. 1 all sources of background are approximated to the measured $\pi^{0}$ sample where 
${\cal{R}}={Y^{a}_{\pi^{0}+h}}/{Y^{n}_{\pi^{0}+h}}$, but, to the extent that this ratio is similar for all 
sources of background and/or the fraction of other sources is small, this approximation 
is valid. 

Figure 2 (right) shows the $z_{T}$ dependence 
of the trigger-normalized fragmentation function for  
p+p and central Au+Au 0-10$\%$. 
Within the NLO calculations for the associated yield with direct $\gamma$ triggers [5], theory
describes the p+p and Au+Au data within errors. 
The away-side yield per direct $\gamma$ is lower than that  
of $\pi^{0}$ for p+p collisions as expected, since  
quarks and gluons are present in different proportions opposite to 
direct $\gamma$ and $\pi^{0}$ triggers, and $\pi^{0}$ results from higher parton initial energy.
It is expected that the away side parton of $\pi^{0}$-ch to travel a longer distance through the medium
and lose more energy due to the bi-colored nature of the gluon. However, 
the results show that the medium has similar effect on 
the multiplicity of the away-side associated charged particles with direct $\gamma$ and $\pi^{0}$ for central Au+Au collisions.

For better quantification, the ratio ($I_{AA}$) of the integrated away-side yield of the associated 
particles per trigger in Au+Au relative to p+p and relative to 
d+Au are shown in Fig.3. 
The similar suppression of the per-trigger away side multiplicity for Au+Au 
with respect to d+Au and p+p indicates the final state effect. 
The suppression of direct $\gamma$ over 
the shown range of $z_{T}$ doesn't indicate a significant dependence of energy loss 
on the parton initial energy and might eliminate the conjectured typical energy loss 
and constant fractional energy loss [6]. 
While the data are well-described at high $z_{T}$ with different theoretical calculations [5,6,7], 
results seem to disfavor the volume emission [5] at low $z_{T}$. 
Despite all the expected differences between the di-hadron and $\gamma$-hadron correlations, 
the value of $I_{AA}^{\gamma h^{\pm}}$ and $I_{AA}^{\pi^{0} h^{\pm}}$ are similar. 

In summary, the energy loss is a final state effect. 
The constant suppression of $I_{AA}^{\gamma h^{\pm}}$ 
with direct photon fractional energy may indicate that the energy loss has no strong dependence on the initial parton energy. 
The similarity of $I_{AA}^{\gamma h^{\pm}}$ and $I_{AA}^{\pi^{0} h^{\pm}}$ indicates either no significant path length dependence of
energy loss or the interplay between different factors blurred the expected difference.
Further extension to low $z_{T}$ may permit the discrimination between different models.

\end{document}